
\magnification=1200
\baselineskip=13pt

\def\sqr#1#2{{\vcenter{\vbox{\hrule height.#2pt
        \hbox{\vrule width.#2pt height#1pt \kern#1pt
          \vrule width.#2pt}
        \hrule height.#2pt}}}}
\def\square{\mathchoice\sqr68\sqr68\sqr{4.2}6\sqr{2.10}6}

\rightline{UR-1386\ \ \ \ \ \ }
\rightline{ER40685-835}

\baselineskip=18pt

\centerline{\bf PATH INTEGRAL SOLUBILITY OF A GENERAL}
\centerline{\bf  TWO-DIMENSIONAL MODEL}

\vskip 1truein

\centerline{Ashok Das}

\centerline{and}

\centerline{Marcelo Hott$^\dagger$}

\centerline{Department of Physics and Astronomy}

\centerline{University of Rochester}

\centerline{Rochester, NY 14627}

\vskip 1 truein

\centerline{\underbar{Abstract}}

\medskip

The solubility of a general two dimensional model, which reduces
 to various models in different limits, is studied within the path
 integral formalism.  Various subtleties and interesting
features are pointed out.

\vskip 2truein

\noindent $^\dagger$On leave
 of absence from UNESP - Campus de
Guaratinguet\'a, P.O. Box 205, CEP : 12.500, Guaratinguet\'a, S.P., Brazil

\vfill\eject

There are a number of $1+1$ dimensional field theoretic models that can be
solved exactly.  The solubility of these models have been studied from
various points of view [1-11].  Normally, these models are formulated in
terms of a fermion field with a vector or axial-vector or a chiral coupling.
 More recently, however, there has been an interest in a model [12,13]
where the fermion has both vector and axial vector couplings of arbitrary
strength.  This model reduces to all other known models in various limits.
In this brief report, we show how this model can be solved in its
generality within the path integral formalism.  We compare our results with
those obtained through a point-splitting regularization [13] and point out
various characteristics of the model.

To begin with, let us consider a fermion in $1+1$ dimension interacting
with an external spin 1 field, described by
$$\eqalign{{\cal L} &= i \overline \psi \gamma^\mu \big( \partial_\mu - i
(1 + r \gamma_5)A_\mu \big) \psi\cr
&= i \overline \psi \gamma^\mu \big( \partial_\mu - i (\eta_{\mu \nu} + r
\epsilon_{\mu \nu} ) A^\nu \big) \psi\cr}\eqno(1)$$
Here \lq$r$' is an arbitrary real parameter and we have used the familiar
identities of $(1+1)$ dimensions in the last line of Eq. (1).  (See ref. 10
for notations, identities and details.) Let us define
$$\widetilde A_\mu = \left( \eta_{\mu \nu} + r \epsilon_{\mu \nu} \right)
A^\nu \eqno(2)$$
and note that in $1+1$ dimensions, we can write
$$\widetilde A_\mu = \partial_\mu \sigma + \epsilon_{\mu \nu} \partial^\nu
\rho \eqno(3)$$
so that the Lagrangian in Eq. (1) takes the form
$${\cal L} = i \overline \psi \gamma^\mu \left( \partial_\mu - i
\widetilde A_\mu \right) \psi = i \overline \psi \gamma^\mu
\left( \partial_\mu - i \partial_\mu \sigma - i \gamma_5 \partial_\mu \rho
\right) \psi \eqno(4)$$
It is clear now that if we define
$$\eqalign{\psi &= e^{i(\sigma + \gamma_5 \rho)} \psi^\prime\cr
\overline \psi &= \overline \psi^\prime e^{-i (\sigma - \gamma_5 \rho)}\cr}
\eqno(5)$$
then the Lagrangian in Eq. (4) reduces to a free theory, namely,
$${\cal L} = i \overline \psi \gamma^\mu \left( \partial_\mu - i
\partial_\mu \sigma - i \gamma_5 \partial_\mu \rho \right) \psi =
i \overline \psi^\prime \gamma^\mu \partial_\mu \psi^\prime \eqno(6)$$
In the path integral formalism, the Jacobian under the field redefinition
in Eq. (5) is nontrivial [14] and we obtain the effective action by
evaluating this Jacobian [9-11].

The evaluation of the Jacobian is straightforward and can be read off from
ref. 10.  However, we would like to emphasize that for the present case, we
can define
$$\widetilde A_\mu^D = \epsilon_{\mu \nu} \widetilde A^\nu \eqno(7)$$
which leads to the identity
$$\gamma^\mu \widetilde A_\mu = \gamma^\mu \left( \eta \widetilde A_\mu
 + \xi \gamma_5 \widetilde A_\mu^D\right)$$
with
$$\eta + \xi = 1 \eqno(8)$$
and one can use the Euclidean Dirac operator
$$\rlap\slash{\rm D}_E = \gamma_\mu \left( \partial_\mu - i
\eta \gamma_\mu \widetilde A_\mu - i \xi \gamma_\mu
\gamma_5 \widetilde A^D_\mu \right) \eqno(9)$$
to evaluate the Jacobian for the change of variables.  We note here that it
is this operator which provides the most general regularization which is
consistent.

For an infinitesimal field redefinition
$$\eqalign{\psi &= e^{i(\epsilon (x) + \gamma_5 \tilde \epsilon
(x))} \psi^\prime\cr
\overline \psi &= \overline \psi ^\prime
e^{-i(\epsilon (x) - \gamma_5 \tilde \epsilon
(x))}\cr}\eqno(10)$$
the Jacobian with the regularization in Eq. (9) can be read off from ref. 10
(simply replace $A_\mu \rightarrow \widetilde A_\mu$ and $A_{5\mu}
\rightarrow \widetilde A^D_\mu$) and has the form
$$J = \exp \left[ - {i \over 2 \pi} \int d^2x_E \left( \eta \epsilon
(x) \epsilon_{\mu \nu} \widetilde F^D_{\mu \nu} + \xi
\widetilde \epsilon (x) \epsilon_{\mu \nu} \widetilde F_{\mu \nu}
\right) \right] \eqno(11)$$
which when rotated to Minkowski space has the form
$$J = \exp \left[ {i \over  \pi} \int d^2x \left( \eta \epsilon
(x) \partial_\mu  \widetilde A^\mu  + \xi
\widetilde \epsilon (x) \partial_\mu  \widetilde A^{\mu, D}
\right) \right] \eqno(12)$$
The anomaly equations for the vector and the axial-vector currents, then,
follow to be $(j^\mu_V = \overline \psi \gamma^\mu \psi,\ j^\mu_A =
\overline \psi \gamma_5 \gamma^\mu \psi)$
$$\eqalign{\partial_\mu j^\mu_V &= {\eta \over \pi} \ \partial_\mu
\widetilde A^\mu = {\eta \over \pi} \ \big( \partial_\mu A^\mu + r
\epsilon^{\mu \nu}
\partial_\mu A_\nu \big)\cr
\noalign{\vskip 4pt}%
\partial_\mu j^\mu_A &= -{\xi \over \pi} \ \partial_\mu
\widetilde A^{\mu, D} = -{\xi \over \pi} \ \big( r \partial_\mu A^\mu +
\epsilon^{\mu \nu}
\partial_\mu A_\nu \big)\cr}\eqno(13)$$
These, of course, reduce to the well known results
 [10] when $r=0$ and we note
that for $j^\mu = j^\mu_V - r j^\mu_A = \overline \psi \gamma^\mu
(1+ r \gamma_5 )\psi$,
$$\partial_\mu j^\mu = {1 \over \pi} \ \left( \eta + \xi r^2 \right)
\partial_\mu A^\mu + {r \over \pi}\
\epsilon^{\mu \nu} \partial_\mu A_\nu \eqno(14)$$
Here $j^\mu$ is the current of our theory in Eq. (1) and we note that it is
anomalous for $r \not= 0$ for any choice of regularization.

The Jacobian for the finite field redefinition in Eq. (5) is again
straightforward following ref. 9 and we obtain
$$J = \exp \left[ - {i \over 2 \pi}\ \int d^2 x \left( \eta_{\mu \sigma} + r
\epsilon_{\mu \sigma}\right) \left( \eta_{\nu \tau} + r
\epsilon_{\nu \tau} \right) A^\sigma
\left( \eta \ {\partial^\mu \partial^\nu \over \square} + \xi
\epsilon^{\lambda \mu} \epsilon^{\rho \nu}
\ {\partial_\lambda \partial_\rho \over \square}\right)
A^\tau \right] \eqno(15)$$
which leads to the effective action
$$\eqalign{Z [A_\mu] &= N \int {\cal D} \overline
\psi {\cal D}
 e^{i \int d^2x {\cal L}}\cr
&= N^\prime \exp \bigg[ - {i \over 2 \pi} \int d^2 x
\big( \eta_{\mu \sigma} + r
\epsilon_{\mu \sigma}\big) \big( \eta_{\nu \tau} + r
\epsilon_{\nu \tau} \big)\cr
&\qquad A^\sigma
\bigg( \eta \ {\partial^\mu \partial^\nu \over \square} + \xi
\epsilon^{\lambda \mu} \epsilon^{\rho \nu}
\ {\partial_\lambda \partial_\rho \over \square}\bigg)
A^\tau \bigg]\cr} \eqno(16)$$
It is straightforward to check that this generating functional yields the
anomaly equation (14).  This can also be checked to coincide with the
result obtained through the point-splitting regularization [13].

Next, let us consider the general model described by [12,13]
$$\eqalign{{\cal L}_{\rm \scriptstyle{TOT}} = &- {1 \over 4}\
\big( \partial_\mu B_\nu - \partial_\nu B_\mu \big) \big(
\partial^\mu B^\nu - \partial^\nu B^\mu \big) + {\mu^2_0
\over 2}\ B_\mu B^\mu\cr
\noalign{\vskip 4pt}%
&+ i \overline \psi \gamma^\mu \big( \partial_\mu - i \big( 1+ r \gamma_5
\big)
\big( A_\mu + e B_\mu \big)\big) \psi + J_\mu B^\mu \cr}\eqno(17)$$
In the generating functional, the fermionic fields can be integrated out to
give the effective action in Eq. (16) with the substitution
$$A_\mu \rightarrow A_\mu + e B_\mu \eqno(18)$$
As a result, we can write
$$Z_{\rm TOT} \left( A_\mu , J_\mu \right) = N \int
{\cal D} B_\mu e^{i S_{\rm eff} (B_\mu , A_\mu , J_\mu )}\eqno(19)$$
where
$$S_{\rm eff} = \int d^2x \left( {1 \over 2}\ B_\mu P^{\mu \nu} B_\nu +
B_\mu Q^\mu + {1 \over 2}\ A_\mu R^{\mu \nu} A_\nu \right) \eqno(20)$$
with
$$\eqalign{P^{\mu \nu} &= \eta^{\mu \nu} \bigg( \square + \mu^2_0 +
{e^2 \over \pi}\ \big( \xi + \eta r^2 \big)\bigg)\cr
\noalign{\vskip 4pt}%
&\qquad -\bigg( 1 + {e^2 \over \pi}\ \big( 1+ r^2\big) \square^{-1} \bigg)
\partial^\mu \partial^\nu - {e^2 r \over \pi}\ \big( \epsilon^{\sigma \mu}
\partial^\nu + \epsilon^{\sigma \nu} \partial^\mu \big) \partial_\sigma
\square^{-1}\cr
\noalign{\vskip 4pt}%
Q^\mu &= J^\mu - {e \over \pi} \  \bigg(
- \big( \xi + \eta r^2 \big) A^\mu\cr
\noalign{\vskip 4pt}%
&\qquad  + \big( 1 + r^2 \big)
{\partial^\mu \partial \cdot A  \over \square}
+ r \big(  \epsilon^{\sigma \mu}
\partial^\nu + \epsilon^{\sigma \nu} \partial^\mu \big) \partial_\sigma
\square^{-1} A_\nu \bigg)\cr
\noalign{\vskip 4pt}%
R^{\mu \nu} &= - {1 \over \pi } \ \bigg( - \big( \xi + \eta r^2 \big) \eta^
{\mu \nu} +
\big( 1+r^2\big) \ {\partial^\mu \partial^\nu
\over \square} + r \big( \epsilon^{\sigma \mu} \partial^\nu +
\epsilon^{\sigma \nu} \partial^\mu \big)
\partial_\sigma \square^{-1}\bigg)\cr}\eqno(21)$$

The action in Eq. (20) is quadratic in $B_\mu$ and hence the generating
functional is easily obtained to be
$$Z_{\rm TOT} \left( A_\mu , J_\mu \right) = N^\prime \exp
\left[ - {i \over 2} \int d^2 x \left( Q^\mu P^{-1}_{\mu \nu} Q^\nu
+ A_\mu R^{\mu \nu} A_\nu \right) \right] \eqno(22)$$
Note that if we define
$$P^{-1}_{\mu \nu} = a \eta_{\mu \nu} + b \partial_\mu
\partial_\nu + c \left( \epsilon_{\sigma \mu} \partial_\nu
 + \epsilon_{\sigma \nu} \partial_\mu \right)
\partial^\sigma \eqno(23)$$
then, from
$$P^{\mu \nu} P^{-1}_{\nu \lambda} = \delta^\mu_\lambda \eqno(24)$$
we can determine
$$\eqalign{a &= {1 \over \square + \mu^2_0 + {e^2 \over \pi}\ (\xi +
 \eta r^2 ) + {(e^2 r/\pi)^2 \over \mu^2_0 + {e^2 \over \pi}\
(\xi + \eta r^2) - {e^2 \over \pi}\ (1 + r^2 )}} =
{1 \over \square + m^2_{\rm phys}}\cr
\noalign{\vskip 4pt}%
b &= {1 \over \mu^2_0 - {e^2 \over \pi}\ (\eta + \xi r^2)} \
{(\square + {e^2 \over \pi}\ (1+ r^2)) \over
(\square + m^2_{\rm phys})}\ {1 \over \square}\cr
\noalign{\vskip 4pt}%
c &= {e^2 r \over \pi (\mu^2_0 - {e^2 \over \pi}\
(\eta + \xi r^2))} \ {1 \over (\square +m^2_{\rm phys})} \
{1 \over \square}\cr}\eqno(25)$$
We can also rewrite
$$m^2_{\rm phys} = \mu^2_0 \ {(1 - {\eta e^2 \over \pi \mu^2_0}\ (1 -
r^2))(1 + {\xi e^2 \over \pi \mu^2_0}\ (1 -r^2)) \over
(1 - {e^2 \over \pi \mu^2_0}\ (\eta + \xi r^2))}\eqno(26)$$
which coincides with the result obtained through the point-splitting
regularization [13].  The propagator for the $B_\mu$-field
 is now seen to be
$$\eqalign{D_{\mu \nu} =\  &{1 \over (\square + m^2_{\rm phys})} \
\bigg[ \eta_{\mu \nu} + {(\square +{e^2 \over \pi}\ (1+ r^2)) \over
(\mu^2_0 - {e^2 \over \pi}\ (\eta +\xi r^2))} \ {\partial^\mu
\partial^\nu \over \square}\cr
\noalign{\vskip 4pt}%
&+ {e^2 r /\pi \over (\mu^2_0 - {e^2 \over \pi}\ (\eta + \xi r^2 ))}
\ \big( \epsilon_{\sigma \mu} \partial_\nu + \epsilon_{\sigma \nu}
\partial_\mu \big) \partial^\sigma \square^{-1} \bigg]\cr}\eqno(27)$$

We end our discussion by noting that the term quadratic in $Q_\mu$ in Eq. (
22) gives rise to a term which is quadratic in $A_\mu$.  (See definition in
Eq. (21).) Consequently, the term quadratic in $A_\mu$ will have a
structure of the form
$${1 \over 2} \int d^2x\ A_\mu R^{\prime \mu \nu} A_\nu \eqno(28)$$
in the exponent of the generating functional.  Consequently, the anomaly
equation derived from this generating functional will differ from that in
Eq. (14).  The reason for this is not hard to understand.  Since both
$A_\mu$ and $B_\mu$ couple to the same current, the one-loop diagrams
contributing to the anomaly will have two parts.

\vskip 2truein

\noindent While Eq. (14) contains the contribution from the first diagram
alone, it is the second diagram which is responsible for the modification
in the anomaly (also in Eq. (28)).

This work is supported in part by the U.S. Department of Energy Grant No.
DE-FG-02-91ER40685.    M.H. would like to thank the
Funda\c c\~ao de Amparo a Pesquisa do Estado de S\~ao
 Paulo for the financial
support.

\vfill\eject

\noindent {\bf \underbar{References}}

\medskip

\item{1.} W. Thirring, Ann. Phys. (N.Y.) {\bf 3},
 91 (1958).

\item{2.} J. Schwinger, Phys. Rev. {\bf 128}, 2425 (1962).

\item{3.} V. Glaser, Nuovo Cimento {\bf 9}, 990 (1958); F. Scarf,
Phys. Rev. {\bf 117}, 868 (1960); T.~Pradhan, Nucl. Phys. {\bf 9},
 124 (1958).

\item{4.} K. Johnson, Nuovo Cimento {\bf 20}, 773 (1961); C.
Sommerfield, Ann. Phys. (N.Y.) {\bf 26}, 1 (1964).

\item{5.} L.S. Brown, Nuovo Cimento {\bf 29}, 617 (1963).

\item{6.} C.R. Hagen, Nuovo Cimento {\bf 51B}, 169 (1967);
Nuovo Cimento {\bf 51A}, 1033 (1967).

\item{7.} C.R. Hagen, Ann. Phys. (N.Y.) {\bf 81}, 67 (1973).

\item{8.} R. Jackiw and R. Rajaraman, Phys. Rev. Lett. {\bf 54},
 1219 (1985).

\item{9.} R. Roskies and F.A. Schaposnik, Phys. Rev. {\bf D23},
 558 (1981).

\item{10.} A. Das and V.S. Mathur, Phys. Rev. {\bf D33}, 489
 (1986).

\item{11.} A. Das, Phys. Rev. Lett. {\bf 55}, 2126 (1985).

\item{12.} A. Bassetto, L. Griguolo and P. Zanca, Phys. Rev. {\bf 50D},
1077 (1994).

\item{13.} C.R. Hagen, University of Rochester preprint UR-1382.

\item{14.} K. Fujikawa, Phys. Rev. Lett. {\bf 42}, 1195 (1979);
 K. Fujikawa, Phys. Rev. {\bf D21}, 2848 (1980); {\it ibid} {\bf D22},
 1499(E) (1980).

\end